\documentclass[
reprint,
amsmath,
amssymb,
aps,prl,superscriptaddress,nofootinbib]{revtex4-1}
\usepackage[pdftex]{graphicx}
\usepackage{wasysym}
\usepackage{amsfonts}
\usepackage{ifsym}

\def\be{\begin{equation}}
\def\ee{\end{equation}}
\def\ba{\begin{eqnarray}}
\def\ea{\end{eqnarray}}

\def\CP1{\mathbb{CP}^1}
\def\SL2C{\mathrm{SL}(2,\mathbb{C})}

\def\Z2{\mathbb{Z}_2}

\def\su2{{SU(2)}}

\def\[{\left[}
\def\]{\right]}

\def\({\left(}
\def\){\right)}
\def\[{\left[}
\def\]{\right]}

\def\<{\langle}
\def\>{\rangle}

\def\i2{\frac{i}{2}}

\def\2F1{\,_2{\rm F}_1}

\usepackage[usenames,dvipsnames]{xcolor}
\usepackage{color}
\usepackage{graphicx}
\usepackage{dcolumn}
\usepackage{bm}
\usepackage{hyperref}

\newcommand{\beq}{\begin{equation}}
\newcommand{\eeq}{\end{equation}}
\newcommand{\beqq}{\begin{equation*}}
\newcommand{\eeqq}{\end{equation*}}
\newcommand\beqa{\begin{eqnarray}}
\newcommand\eeqa{\end{eqnarray}}
\newcommand\beqaa{\begin{eqnarray*}}
\newcommand\eeqaa{\end{eqnarray*}}
\newcommand\bea{\begin{array}}
\newcommand\eea{\end{array}}

\begin{document}


\title{Scattering of Massless Particles in Arbitrary Dimension}


\author{Freddy Cachazo}
\email{fcachazo@perimeterinstitute.ca}
\affiliation{Perimeter Institute for Theoretical Physics, Waterloo, ON N2L 2Y5, Canada}
\author{Song He}
\email{she@perimeterinstitute.ca}
\affiliation{Perimeter Institute for Theoretical Physics, Waterloo, ON N2L 2Y5, Canada}
\affiliation{School of Natural Sciences, Institute for Advanced Study, Princeton, NJ 08540, USA}
\author{Ellis Ye Yuan}
\email{yyuan@perimeterinstitute.ca}
\affiliation{Perimeter Institute for Theoretical Physics, Waterloo, ON N2L 2Y5, Canada}
\affiliation{Physics Department, University of Waterloo, Waterloo, ON N2L 3G1, Canada}


\date{\today}

\begin{abstract}
We present a compact formula for the complete tree-level S-matrix of pure Yang-Mills and gravity theories in arbitrary spacetime dimension. The new formula for the scattering of $n$ particles is given by an integral over the position of $n$ points on a sphere restricted to satisfy a dimension-independent set of equations. The integrand is constructed using the reduced Pfaffian of a $2n\times 2n$ matrix, $\Psi$, that depends on momenta and polarization vectors. In its simplest form, the gravity integrand is a reduced determinant which is the square of the Pfaffian in the Yang-Mills integrand. Gauge invariance is completely manifest as it follows from a simple property of the Pfaffian.
\end{abstract}


\maketitle

\section{Introduction}

In a recent work \cite{Cachazo:2013gna}, we pointed out the existence of equations connecting the space of kinematic invariants of $n$ massless particles in any dimension and that of the position of $n$ points on a sphere. The equations are given by
\be
\sum_{b\neq a} \frac{k_a\cdot k_b}{\sigma_{a}-\sigma_{b}} = 0 \quad {\rm for} \quad a\in \{ 1,2,\ldots , n\}
\label{scatt}\ee
where $\sigma_c$ is the position of the $c^{\rm th}$ puncture. Motivated by some remarkable properties of these equations, namely their connection to general kinematic invariants and Kawai-Lewellen-Tye (KLT) orthogonality\footnote{The KLT orthogonality proved in \cite{Cachazo:2013gna} is the fact that ``Parke-Taylor" vectors evaluated on distinct solutions of the scattering equations are orthogonal with respect to the kernel of the field-theory KLT relations, which relate Yang-Mills and gravity amplitudes~\cite{Kawai:1985xq}.}, it was proposed that they are the backbone of the tree-level S-matrix of massless particles in any dimension and were called the {\it scattering equations}.

These equations are invariant under $\SL2C$ transformations when external vectors satisfy momentum conservation. The scattering equations were first discovered, in the context of field theory amplitudes, by one of the authors\footnote{The equations in \cite{Cachazo:2012uq} and those in \eqref{scatt} are related by the $\SL2C$ transformation $S = \left(
                                             \begin{array}{cc}
                                               0 & 1 \\
                                               -1 & 0 \\
                                             \end{array}
                                           \right)$.
} in the study of the fundamental Bern-Carrasco-Johansson (BCJ) relations~\cite{Bern:2008qj} in four dimensions \cite{Cachazo:2012uq}. The validity of the BCJ relations for gauge-theoretical amplitudes in any dimension~\cite{Bern:2008qj, BjerrumBohr:2009rd, Stieberger:2009hq} also provides an important piece of evidence for the universal relevance of the scattering equations.

In \cite{Cachazo:2013gna} we also proposed the existence of formulas for the complete S-matrix of Yang-Mills and gravity theories in any dimension. In this Letter we provide the explicit construction of such formulas.

\section{Preliminaries}

The first step towards the construction of formulas in any spacetime dimension is that of the measure. Given that only $n-3$ of the $n$ scattering equations are linearly independent one has to find a way of imposing their support in a permutation invariant manner. This is achieved by noticing that
\be
\prod_a {}'\delta(\sum_{b\neq a} \frac{k_{a}\cdot k_{b}}{\sigma_{a b}}) \equiv \sigma_{ij}\sigma_{jk}\sigma_{ki}\prod_{a\neq i,j,k}\delta(\sum_{b\neq a} \frac{k_{a}\cdot k_{b}}{\sigma_{a b}})
\ee
is independent of the choice $\{i,j,k\}$ and hence permutation invariant. Here and in the rest of this paper $\sigma_{ab} = \sigma_a-\sigma_b$.

Let us denote the $n$-gluon partial amplitude with the canonical ordering $1,2,\ldots, n$ as $A_n$, and the $n$-graviton amplitude as $M_n$. It is now natural to propose the following formulations of their S-matrices,
\ba \!\!\!\! A_n\! &=&\! \int \frac{d\,^n\sigma}{\textrm{vol}\,\SL2C} \prod_a {}'\delta(\sum_{b\neq a} \frac{k_{a}\cdot k_{b}}{\sigma_{a b}}) \frac{E_n(\{k,\epsilon,
\sigma\})}{\sigma_{12}\ldots\sigma_{n1}},~\label{YMgen}\\ \!\!\!\! M_n\! &=&\! \int\frac{d\,^n\sigma}{\textrm{vol}\,\SL2C} \prod_a {}'\delta(\sum_{b\neq a} \frac{k_{a}\cdot k_{b}}{\sigma_{a b}}) E^2_n(\{k,\epsilon, \sigma\}) ~\label{gravitygen}
\ea
where $E_n(\{k,\epsilon, \sigma\})$ is a permutation invariant function of $\sigma_a$, momenta $k_a^\mu$ and polarization vectors $\epsilon_a^\mu$. Note that $\SL2C$ invariance of the integrand constrains $E_n$: under an $\SL2C$ transformation, $\sigma_a\to \frac{A\sigma_a+B}{C\sigma_a+D}$, $E$ must transform as
\be\label{SL2C}
E_n(\{k,\epsilon,
\sigma\})\to E_n(\{k,\epsilon,
\sigma\}) \prod_{a=1}^n(C\sigma_a+D)^2.
\ee
It is also natural to expect that $E_n$ should be gauge invariant for each solution of the scattering equations.

At this point it is worth to spell out how the measure is computed in practice which uncovers a beautiful relation to a matrix found previously in the literature and hence shows its permutation invariance. Consider the object
\be
\int \frac{d\,^n\sigma}{\textrm{vol}\,\SL2C} \prod_a {}'\delta(\sum_{b\neq a} \frac{k_{a}\cdot k_{b}}{\sigma_{a b}}) ~\bullet
\label{fp}\ee
where $\bullet$ represents either the integrand of Yang-Mills or that of gravity. Using a Fadeev-Popov procedure to gauge fix the $\SL2C$ redundancy and hence fix the value of, say, $\sigma_p,\sigma_q,\sigma_r$ one finds that \eqref{fp} becomes
\be
\int\!\!\!\prod_{c\neq p,q,r}\!\!\! d\sigma_c \, (\sigma_{pq}\sigma_{qr}\sigma_{rp})(\sigma_{ij}\sigma_{jk}\sigma_{ki}) \!\!\prod_{a\neq i,j,k}\delta(\sum_{b\neq a} \frac{k_{a}\cdot k_{b}}{\sigma_{a b}})~\bullet
\ee
The delta functions completely localize all integrals. As proven in \cite{Cachazo:2013gna} the scattering equations have $(n-3)!$ solutions and the answer is obtained by evaluating a Jacobian and the integrand on them. The Jacobian can be computed by starting with an $n\times n$ matrix $\Phi$ defined by
\be
\Phi_{ab} = \begin{cases}
\displaystyle ~~\frac{k_{a}\cdot k_{b}}{(\sigma_a-\sigma_b)^2} & a\neq b,\\
\displaystyle -\sum_{c\neq a}\frac{k_{a}\cdot k_{c}}{(\sigma_a-\sigma_c)^2} & a=b.
\end{cases}
\ee
The fact that the delta functions exclude $\{ i,j,k\}$ means that we have to delete those rows from $\Phi$ while having fixed the values of $\sigma_p,\sigma_q,\sigma_r$ means that we have to delete columns $\{ p,q,r\}$. Let us denote the corresponding minor by $|\Phi|^{ijk}_{pqr}$. This minor is the Jacobian we are after. The answer is then
\be
\sum_{\{\sigma\}\in \text{ solutions}}\frac{(\sigma_{pq}\sigma_{qr}\sigma_{rp})(\sigma_{ij}\sigma_{jk}\sigma_{ki})}{|\Phi|^{ijk}_{pqr}}~\bullet
\label{perf1}\ee
Precisely the combination that appears in this equation is what was called ${\det}'\Phi$ by Cachazo and Geyer in \cite{Cachazo:2012da} (inspired by a remarkable formula for MHV gravity amplitudes found by Hodges in \cite{Hodges:2012ym}) and which is known to be completely permutation invariant, i.e., independent of the choices made in selecting $\{ i,j,k\}$ and $\{ p,q,r\}$. More explicitly,
\be
{\det}'\Phi \equiv  \frac{|\Phi|^{ijk}_{pqr}}{(\sigma_{pq}\sigma_{qr}\sigma_{rp})(\sigma_{ij}\sigma_{jk}\sigma_{ki})}.
\label{perf2}\ee

\section{Explicit Form of $E_n(\{k,\epsilon,\sigma\})$}

In order to present the explicit form of $E_n(\{k,\epsilon,\sigma\})$, first
define the following $2n\times 2n$ antisymmetric matrix
\be\label{Psi}
\Psi = \left(
         \begin{array}{cc}
           A &  -C^{\rm T} \\
           C & B \\
         \end{array}
       \right)
\ee
where $A$, $B$ and $C$ are $n\times n$ matrices. The first two matrices have components
\be
A_{ab} = \begin{cases} \displaystyle \frac{k_a\cdot k_b}{\sigma_{a}-\sigma_{b}} & a\neq b,\\
\displaystyle \quad ~~ 0 & a=b,\end{cases} \quad\quad B_{ab} = \begin{cases} \displaystyle \frac{\epsilon_a\cdot \epsilon_b}{\sigma_{a}-\sigma_{b}} & a\neq b,\\
\displaystyle \quad ~~ 0 & a=b\end{cases}
\label{ABmatrix}
\ee
while the third is given by
\be
C_{ab} = \begin{cases} \displaystyle \frac{\epsilon_a\cdot k_b}{\sigma_{a}-\sigma_{b}} &\quad a\neq b,\\
\displaystyle -\sum_{c\neq a}\frac{\epsilon_a\cdot k_c}{\sigma_{a}-\sigma_{c}} &\quad a=b.\end{cases}
\ee

The first important observation is that while the Pfaffian of $\Psi$ is zero, removing rows $i,j$ and columns $i,j$ with $1\leq i<j\leq n$ gives rise to a new matrix $\Psi_{ij}^{ij}$ with non-zero Pfaffian and such that
\be
{\rm Pf}'\Psi \equiv 2\frac{(-1)^{i+j}}{(\sigma_i-\sigma_j)} {\rm Pf}(\Psi_{ij}^{ij})
\ee
is independent of the choice of $i$ and $j$. We call ${\rm Pf}'\Psi$ the reduced Pfaffian of $\Psi$.

The Pfaffian of $\Psi$ vanishes because its first $n$ rows (or columns) are linearly dependent: actually the $n\times 2n$ matrix $(A,-C^{\rm T})$ has two null vectors, $(1,\ldots,1)$ and $(\sigma_1,\ldots,\sigma_n)$, thus $({\rm Pf}\Psi)^2=\det \Psi=0$. Now we turn to the proof that the reduced Pfaffian is invariant under permutations of {\it particle labels}. First note that simultaneous interchanges of two columns and rows change the sign of the Pfaffian. When exchanging two particle labels $a,b$ which are different from $i,j$, we must exchange rows and columns $a,b$ and also exchange $a{+}n, b{+}n$; when exchanging particle labels $i,j$, we only exchange $i{+}n, j{+}n$ in $\Psi^{ij}_{ij}$ and the additional minus sign cancels with the minus sign from the prefactor in the definition of ${\rm Pf}'\Psi$. Hence, in both cases the reduced Pfaffian is invariant. Therefore, to prove permutation invariance, it suffices to prove that the reduced Pfaffian obtained from removing columns and rows $1,2$ and that from removing $1,3$ are identical.

We multiply the first row and column of $\Psi^{12}_{12}$ by $\sigma_{13}$, and the first row and column of $\Psi^{13}_{13}$ by $\sigma_{12}$, and obtain two matrices we call $\Psi'{}^{12}_{12}$ and $\Psi'{}^{13}_{13}$. Next we take a multiple of the $(i{-}2)^{\rm th}$ row and column of $\Psi'{}^{12}_{12}$ by $\sigma_{1i}$ and add all the multiples to the first row and column respectively, for $i=4,\ldots,n$; in this way we obtain a new matrix $\Psi''{}^{12}_{12}$, and similarly we have $\Psi''{}^{13}_{13}$, whose Pfaffians are related to the original ones by ${\rm Pf}\Psi''{}^{12}_{12}=\sigma_{13}{\rm Pf}\Psi^{12}_{12}$, ${\rm Pf}\Psi''{}^{13}_{13}=\sigma_{12}{\rm Pf}\Psi^{13}_{13}$.
By the scattering equations, it is straightforward to show that the first row and column of $\Psi''{}^{12}_{12}$ only differ from the first row and column of $\Psi''{}^{13}_{13}$ by a minus sign; note that other columns and rows of the two new matrices are identical, thus $\frac{{\rm Pf}\Psi^{12}_{12}}{\sigma_{12}}=-\frac{{\rm Pf}\Psi^{13}_{13}}{\sigma_{13}}$. We conclude that the reduced Pfaffian ${\rm Pf}'\Psi$ is permutation invariant with respect to the particle labels.

Now we are ready to write down the proposal
\be
E_n(\{k,\epsilon,
\sigma\}) ={\rm Pf}'\Psi(k, \epsilon,\sigma ).
\label{eform}\ee

Combining this proposal for $E_n$ with the general formula \eqref{YMgen} and \eqref{gravitygen} gives the main results of this paper: A formula for the tree-level S-matrix of Yang-Mills in any dimension
%
%
%
\be\label{YM}
A_n=\frac{1}{\textrm{vol}\,\SL2C}\int \frac{d\,^n\sigma}{\sigma_{12}\ldots\sigma_{n1}} \prod_a {}'\delta(\sum_{b\neq a} \frac{k_a\cdot k_b}{\sigma_{a b}})\,{\rm Pf}'\Psi.
\ee
And, using the KLT construction in the form discussed in \cite{Cachazo:2013gna, Cachazo:2012da} and the KLT orthogonality proven in \cite{Cachazo:2013gna}, a formula for the tree-level S-matrix of gravity
%
%
%
\be
M_n= \frac{1}{\textrm{vol}\,\SL2C}\int d\,^n\sigma \prod_a {}'\delta(\sum_{b\neq a} \frac{k_a\cdot k_b}{\sigma_{a b}})\,{\rm Pf}'\Psi\,{\rm Pf}'\tilde\Psi.
\ee
Here $\tilde\Psi$ is taken to mean $\Psi(k, \tilde\epsilon,\sigma )$ and where $\tilde\epsilon_a$ represents the same physical polarization as $\epsilon_a$. In its simplest form, one can choose $\tilde\epsilon_a =\epsilon_a$ and obtain
\be
M_n= \frac{1}{\textrm{vol}\,\SL2C}\int d\,^n\sigma \prod_a {}'\delta(\sum_{b\neq a} \frac{k_a\cdot k_b}{\sigma_{a b}})\,{\rm det}'\Psi.
\ee
where ${\rm det}'\Psi$ is defined as $4{\rm det}\Psi^{ij}_{ij}/\sigma^2_{ij}$.

Finally, it is worth to also write both formulas in a form where all the integrals have been performed using \eqref{perf1} and \eqref{perf2}
\be
A_n=\sum_{\{\sigma\}\in\text{ solutions}}\frac{1}{\sigma_{12}\ldots\sigma_{n1}}\frac{{\rm Pf}'\Psi(k, \epsilon,\sigma )}{{\det}'\Phi}
\label{each1}\ee
and
\be
M_n=\sum_{\{\sigma\}\in\text{ solutions}}\frac{{\rm det}'\Psi(k, \epsilon,\sigma )}{{\det}'\Phi}.
\label{each2}\ee

\section{Properties and Checks}

Simple properties such as multilinearity in polarization vectors, $\SL2C$ invariance and its mass dimension are easy to check by using the expansion of the Pfaffian or its recursion relation (analogous to those of the determinant). Gauge invariance, as the statement that the amplitude vanishes when any $\epsilon^\mu_a$ is replaced by a multiple of $k_a^\mu$, is obvious since two columns of the matrix $\Psi$ and hence of $\Psi^{ij}_{ij}$ become multiples of each other under the replacement. More explicitly, assume that $\epsilon_i^\mu$ is replaced by $k_i^\mu$, then it can be easily seen that columns $i$ and $i+n$ of $\Psi$ become identical after realizing that
\be
C_{ii} = -\sum_{c\neq i}\frac{\epsilon_i\cdot k_c}{\sigma_i-\sigma_c} \rightarrow -\sum_{c\neq i}\frac{k_i\cdot k_c}{\sigma_i-\sigma_c} = 0
\ee
by the scattering equations. The last property which is manifest is the behavior under soft limits. As discussed in detail in \cite{Cachazo:2013gna}, when we take $k_n\to 0$, $n-1$ of the scattering equations become identical to those of a system with $n-1$ particles. The last equation
\be
\sum_{b\neq n}\frac{k_n\cdot k_b}{\sigma_n-\sigma_b} = 0
\ee
becomes a polynomial for $\sigma_n$ of degree $n-3$ (due to momentum conservation). In this discussion we focus on Yang-Mills amplitudes. It is convenient to compute ${\rm Pf}'\Psi$ using $\Psi^{ij}_{ij}$ with $i\neq n, j\neq n$. The Pfaffian of a $2m\times 2m$ matrix $E$ satisfies a recursion relation of the form ${\rm Pf}(E) = \sum_{q=1}^{2m}(-1)^q e_{pq}{\rm Pf}(E_{pq}^{pq})$. Using this formula to expand ${\rm Pf}\Psi^{ij}_{ij}$ setting $p=n$ one finds that in the soft limit only one term contributes and gives
\be
{\rm Pf}\Psi^{ij}_{ij} \to C_{nn}{\rm Pf}\Psi^{ijn(2n)}_{ijn(2n)}.
\ee
Very nicely, ${\rm Pf}\Psi^{ijn(2n)}_{ijn(2n)}$ is independent of $k_n$ and $\epsilon_n$ and leads to ${\rm Pf}'\Psi_{n-1}$, i.e., the reduced Pfaffian for $n-1$ particles. Using the explicit formula \eqref{YM} in the soft limit one finds
\be
A_n\to \sum_{i=1}^{(n-4)!}\oint_{\Gamma} d\sigma_n\frac{\sum_{a\neq n}\frac{\epsilon_n\cdot k_a}{\sigma_{na}}}{\sum_{a\neq n}\frac{k_n\cdot k_a}{\sigma_{na}}}\frac{\sigma_{n-1,1}}{\sigma_{n-1,n}\sigma_{n,1}}{\cal I}_{n-1}^{(i)}
\ee
where ${\cal I}_{n-1}^{(i)}$ are the terms in the expansion of \eqref{each1} for $A_{n-1}$ and all $\sigma_a$'s with $a\in \{1,2,\ldots ,n-1\}$ are taken to be evaluated on the $i^{\rm th}$ solution. Also, the contour $\Gamma$ is defined to encircle the $n-3$ zeroes of the first factor in the denominator\footnote{Since $\sigma_a$'s are taken to be complex numbers in this paper, the delta functions imposing the scattering equations are in fact poles and all our integrals are contour integrals.}. Using the residue theorem one finds that there is no contribution at infinity and only two poles have non-vanishing residue. These are at $\sigma_{n}=\sigma_{n-1}$ and at $\sigma_{n}=\sigma_{1}$. The residues are trivial to compute as only one term from the sum in the numerator and one from that in the denominator contribute giving rise to
\be
A_n\to \left(\frac{\epsilon_n\cdot k_{n-1}}{k_n\cdot k_{n-1}} + \frac{\epsilon_n\cdot k_{1}}{k_n\cdot k_{1}}\right) A_{n-1}
\ee
which is the correct soft behavior~\cite{Weinberg:1964ew,Weinberg:1965nx}. A completely analogous computation gives the correct soft behavior for gravity as well. Factorization of the amplitude on physical poles is a more involved computation and details are provided in \cite{webpage}.

We have also performed some non-trivial checks such as the agreement of our formula for gluons with formulas available in the literature for three-, four- and five-particle scattering in any dimension (see e.g.~\cite{Green:1987sp} for $n=3,4$, and~\cite{Medina:2002nk} for $n=5$). The case with five particles is the most interesting one as the scattering equations in dimensions greater than four do not factor and the two solutions come from an irreducible quadratic equation. This is the first case that our formula clearly computes the amplitudes in a novel way. We also performed numerical checks that when evaluated in four dimensional kinematics, our formula reproduces all amplitudes with $n\leq 8$ and in all possible helicity sectors (including the all plus and all but one plus).

\section{Discussions}

We have presented a formula for the complete tree level S-matrix of gluons and gravitons in any spacetime dimension. While formulas in dimensions less than ten could exploit the presence of supersymmetry in defining an on-shell superspace, such as the Witten-RSV formula does in four dimensions \cite{Witten:2003nn,Roiban:2004yf}, our formula necessarily depends on polarization vectors as it is also valid in dimensions where supersymmetry does not exist. Any formula which contains polarization vectors must satisfy the constraint that it vanishes when any polarization vector is replaced by a multiple of its momentum vector. What we have found in this work is that there exist a very compact formula in which gauge invariance is actually a simple property of its intrinsic structure and indeed it was the main clue for its derivation.

As discussed in~\cite{Stieberger:2013hza,Stieberger:2013nha} and~\cite{Cachazo:2013gna}, there are compact formulas for string amplitudes in terms of Yang-Mills/gravity amplitudes, and our proposal here also provides a simple representation of string amplitudes in terms of polarization vectors. In relation to string amplitudes, it is important to mention an intriguing connection to their high energy scattering limit. In the work of Gross and Mende \cite{Gross:1987ar}, the scattering equations appear as the conditions imposed by the saddle point evaluation of the string amplitude. It is tempting to suggest that this is more than a coincidence.

Scattering amplitudes of gluons and gravitons can also be obtained in any dimension by using a BCFW recursion relation as proven in \cite{ArkaniHamed:2008yf}. It would be interesting to find a connection between our formula and the BCFW construction of \cite{ArkaniHamed:2008yf} perhaps through some contour deformation argument.

Finally, also worth mentioning is that all $(n-3)!$ solutions of the scattering equations give rise to gauge invariant contributions. Moreover, under factorization limits each term in \eqref{each1} and \eqref{each2} either develops a pole and `factors' or it remains finite. This is reminiscent of the behavior of partial amplitudes in Yang-Mills theory where a decomposition of the full amplitude is made in parts that do not exhibit all factorization channels. Adding the fact that in dimensions greater than four the $(n-3)!$ do not split into sectors, it is natural to suggest that each solution is a `partial amplitude'. In Yang-Mills, this decomposition is in addition to the usual color ordering one while in gravity it is all there is. It would be fascinating to fully uncover the physical meaning of this new decomposition.


{\it Acknowledgements:}
This work is supported by Perimeter Institute for Theoretical Physics. Research at Perimeter Institute is supported by the Government of Canada through Industry Canada and by the Province of Ontario through the Ministry of Research \& Innovation.

\end{document}